\documentclass[11pt]{amsart}
\usepackage{graphicx,amssymb,amsmath,amsthm}
\usepackage{cite}
\usepackage{geometry}
\usepackage{enumitem}
\usepackage{amsfonts}

\usepackage{hyperref} 
\hypersetup{
    colorlinks=true, 
    linkcolor=blue,  
    urlcolor=cyan,   
    filecolor=magenta, 
    citecolor=green  
}
\newtheorem{definition}{Definition}[section]
\newtheorem{problem}{Problem}[section]
\newtheorem{example}{Example}[section]
\newtheorem{question}{Problem}[section]
\newtheorem{corollary}[definition]{Corollary}
\newtheorem{theorem}[definition]{Theorem}
\newtheorem{lemma}[definition]{Lemma}
\newtheorem{remark}[definition]{Remark}

\usepackage{comment}
\usepackage{algorithm}  
\usepackage{algorithmic} 

\newcommand{\shsp}{\hspace{1em}}

\usepackage{tikz}

\usetikzlibrary{arrows}
\tikzstyle{block}=[draw opacity=0.7,line width=1.4cm]

\newcommand{\vu}{\mathbf{u}}

\newcommand{\R}{\mathbb{R}}
\newcommand{\C}{\mathbb{C}}

\newcommand{\inner}[1]{\langle #1 \rangle}
\renewcommand{\H}{{\mathbb F}}
\newcommand{\F}{{\mathbb F}}
\newcommand{\Z}{\mathbb{Z}}

\newcommand{\innerp}[1]{\langle {#1} \rangle}

\newcommand{\Tr}{{\rm Tr}}

\newcommand{\abs}[1]{\lvert#1\rvert}

\newcommand{\ML}{{\mathcal M}}
\newcommand{\cM}{{\mathcal M}}
\newcommand{\cS}{{\mathcal S}}

\newcommand{\rank}{{\rm rank}}

\newcommand{\vx}{{\mathbf x}}
\newcommand{\vz}{{\mathbf z}}
\newcommand{\vy}{{\mathbf y}}

\newcommand{\va}{{\mathbf a}}
\newcommand{\vA}{{\mathbf A}}

\newcommand{\VA}{{\mathbf A}}

\newcommand{\wmod}[1]{\mbox{~(mod~$#1$)}}
\newcommand{\m}{\mathfrak{m}}

\newcommand{\A}{{\mathcal A}}

\newcommand{\MM}{\mathbf M}
\newcommand{\lra}{{\longrightarrow}}

\newcommand{\argmin}[1]{\mathop{\rm argmin}}
\begin{document}
\baselineskip 20pt
	\title{Signal Recovery  on Algebraic Varieties Using Linear Samples}
	
	\author{Zhiqiang Xu}
	\thanks{ Zhiqiang Xu is supported  by
the National Science Fund for Distinguished Young Scholars (12025108), NSFC (12471361, 12021001, 12288201) and National Key R\&D Program of China (2023YFA1009401). }
\address{State Key Laboratory of Mathematical Sciences, Academy of Mathematics and Systems Science, Chinese Academy of Sciences, Beijing 100190, China;   2. School of Mathematical Sciences, University of Chinese Academy of Sciences, Beijing 100049, China.}
\email{xuzq@lsec.cc.ac.cn}
	\maketitle

	\begin{abstract}
The recovery of an unknown signal from its linear measurements is a fundamental problem spanning numerous scientific and engineering disciplines. Commonly, prior knowledge suggests that the underlying signal resides within a known algebraic variety. This context naturally leads to a  question: what is the minimum number of measurements required to uniquely recover any signal belonging to such an algebraic variety?
 In this survey paper, we introduce a method that leverages  tools from algebraic geometry to address this  question. We then demonstrate the utility of this approach by applying it to two problems: phase retrieval and low-rank matrix recovery. 
We also highlight several open problems, which could serve as a basis for future investigations in this field.
	\end{abstract}
\section{Introduction}

\subsection{Illustrative Example: A Simplified Case Study}
Let us consider the problem of recovering an arbitrary vector $\vx$, given that $\vx$ is an element of the two-dimensional real vector space $\R^2$. The information we gather about $\vx$ is the linear samples $y_j=\innerp{\va_j,\vx}, j=1,\ldots,m$.
It is a well-established principle that to uniquely recover a vector $\vx \in \R^2$, the number of measurements, denoted by $m$, must be at least $2$. In the specific case where $m = 2$, to successfully recover $\vx \in \R^2$  from the measurements $y_1 = \innerp{\va_1,\vx}$ and $y_2 = \innerp{\va_2,\vx}$, it is necessary to ensure that the determinant condition $a_{1,1}a_{2,2} - a_{1,2}a_{2,1} \neq 0$ is satisfied, where $\va_1=(a_{1,1}, a_{1,2})^\top\in \R^2$ and $\va_2=(a_{2,1},a_{2,2})^\top\in \R^2$. 
Consider the coefficient matrix $\vA=(\va_1,\va_2)$  as a point in $\R^4$, represented as $(a_{1,1},a_{1,2},a_{2,1},a_{2,2})\in \R^4$. The determinant condition is then equivalent to
\[
\vA\notin V:=\{(x,y,z,w)\in \R^4: xw-yz=0\}.
\]

Let us now proceed under the assumption that $\vx$ is known a priori to belong to the set $W$, where
$W := \{(x_1,x_2)\in \R^2: x_1^2-4x_2^2=0\}$.  Then we can recover  $\vx\in W\subset \R^2$ from one linear sample  $\innerp{\va_0,\vx}=4$ where $\va_0=(1,2)^\top$. 
Indeed, a straightforward calculation reveals that $\vx=(2,1)^\top$ is the unique solution satisfying both the linear measurement and the set constraint.
The reason we can use only one linear sample to recover a two-dimensional vector is that we have prior knowledge that $\mathbf{x}$ belongs to  $W$.

Extending from the illustrative example, we now formulate the general problem as follows:
 
\begin{question}\label{que1}
 {Let $W \subset \C^d$
 be an algebraic variety, which is the zero locus of a finite collection of polynomials in $\C[\vx]$.
   We seek to determine the minimal integer $m$ for which there exists a linear sampling operator $\vA: \F^d \to \F^m$
such that any $\vx \in W$ can be uniquely recovered from its image $\vA\vx$, where $\F\in \{\R,\C\}$. In other words, what is the minimal number of linear measurements required to ensure injective sampling on $W$?}
 \end{question}

\subsection{Problem formulation}

 To generalize Problem \ref{que1}, we introduce the following definitions:
Let $L: \mathbb{C}^d \times \mathbb{C}^d \rightarrow \mathbb{C}$ denote a bilinear function. We define a linear sample of $\mathbf{x} \in \mathbb{C}^d$ as $y = L(\mathbf{a}, \mathbf{x})$, where $\mathbf{a} \in \mathbb{C}^d$ is a known vector. For $j = 1, \ldots, m$, let $\mathbf{a}_j \in V_j$ be sampling vectors, where $V_j \subset \mathbb{C}^d$ is an algebraic variety. We set $\vA:=(\va_1,\ldots,\va_m)^\top\in \C^{m\times d}$. For $\mathbf{x} \in W$, we define $L_\vA: \mathbb{C}^d \rightarrow \mathbb{C}^m$ to represent the linear samples of $\mathbf{x}$ as follows:
\[
L_\vA(\mathbf{x}) = (L(\mathbf{a}_1, \mathbf{x}), L(\mathbf{a}_2, \mathbf{x}), \ldots, L(\mathbf{a}_m, \mathbf{x})) \in \mathbb{C}^m.
\]
This formulation encapsulates the process of obtaining $m$ linear samples of $\mathbf{x}$ using the bilinear function $L$ and the sampling vectors $\mathbf{a}_j\in V_j, j=1,\ldots,m$. 
We say $L_\vA$ is {\em injective} on $W \subseteq \mathbb{C}^d$ if $L_\vA(\vx) = L_\vA(\vy)$ implies $\vx = \vy$ for all $\vx, \vy \in W$. Note that $L_\vA(\vx)=L_\vA(\vy)$ is equivalent to 
$L_\vA(\vx-\vy)=0$.  Hence, $L_\vA$ is injective on $W$ if and only if $L_\vA(\vx)=0$ for $\vx\in W-W$ just has zero solution where $W-W:=\{\vx-\vy: \vx,\vy\in W\}$.

  It is worth noting that even when $W$ is an algebraic variety, $W-W$ may not necessarily be an algebraic variety itself. In this paper, we focus our attention on the case where both $W$ and $W-W$ are algebraic varieties. 
   For convenience, we will continue to use the notation $W$ instead of $W - W$.
We can now generalize Problem \ref{que1} and formulate it as follows:
 \begin{question}\label{que2}
Let $W \subset \mathbb{C}^d$ and $V_j\subset \mathbb{C}^d, j=1,\ldots,m$ be algebraic varieties.
 We seek to determine the minimal value of $m$ for which there exists a matrix $\vA = (\mathbf{a}_1,\ldots,\mathbf{a}_m)^\top \in \mathbb{C}^{m \times d}$, with $\mathbf{a}_j \in V_j$ for $j = 1,\ldots,m$, such that the following condition is satisfied:
 \begin{center}
 For $\mathbf{x} \in W$, $L_\vA(\mathbf{x}) = \mathbf{0}$ implies $\mathbf{x} = \mathbf{0}$.
  \end{center}
 \end{question}
 
 \subsection{Organization of the Paper} This paper is organized as follows: Section 2 introduces two concepts, algebraic variety and admissible condition, which are then elucidated with illustrative examples. Following this, 
 Section 3 presents a comprehensive overview of the general framework of algebraic variety methods, as developed in previous studies. Sections 4 and 5 then explore concrete applications of this framework, building upon the theoretical foundations outlined in Section 3. Specifically, Section 4 examines the application of these methods to generalized phase retrieval. Section 5 extends the discussion to  low-rank matrix recovery.
 
 \begin{remark}
 The application of algebraic varieties to the minimal measurement problem in phase retrieval was first introduced in \cite{BCE06}. This  approach was subsequently refined and  developed in \cite{CEHV15,Edidin}, yielding more comprehensive results concerning the minimum measurement number  required for phase retrieval. 
  Beyond phase retrieval, the algebraic variety method was also  employed in \cite{Xu15} to study the minimal measurement problem for low-rank matrix recovery. Subsequently, in \cite{WangXu}, the method was  generalized and developed, extending its applicability beyond particular problem instances.
 \end{remark}
  
\section{Preliminaries }

\subsection{Algebraic variety}
Throughout this paper, we define an algebraic variety $W \subset \C^d$
  as the zero locus of a finite collection of polynomials in $\C[\vx]$. More precisely:
\[
W = \{\vx \in \C^d : p_1(\vx) = \cdots = p_k(\vx) = 0\},
\]
where $p_1, \ldots, p_k \in \C[\vx]$ are polynomials in $d$ complex variables.
The dimension of an algebraic variety $W$, denoted by $\dim(W)$, is a fundamental quantity intrinsically associated with $W$. For a precise definition of $\dim(W)$, we refer the reader to Definition 2.3 in \cite{dim}.

\begin{example}

We denote by $\cM_{d,r}\subset \C^{d\times d}$ the set of $d\times d$ complex matrices with rank less than or equal to $r$, formally defined as:
\begin{equation}\label{eq:cmdr}
\cM_{d,r}\,\,:=\,\, \{Q\in \C^{d\times d}: {\rm rank}(Q)\leq r\}.
\end{equation}
 It is worth noting that the condition $\text{rank}(Q) \leq r$ is equivalent to the vanishing of all $(r+1) \times (r+1)$ minors of $Q$. These minors are homogeneous polynomials in the entries of $Q$. As a result, $\mathcal{M}_{d,r}$
  forms an algebraic variety in $\mathbb{C}^{d^2}$. According to \cite[Prop. 12.2]{dim}, the dimension of this variety is given by $\dim(\cM_{d,r})=2dr-r^2$.

\end{example}

Let \( W_{\mathbb{R}} := W \cap \mathbb{R}^d \), which consists of all real points in \( W \). We denote by \( \dim_{\mathbb{R}}(W_{\mathbb{R}}) \) the real dimension of this set. When \( W_{\mathbb{R}} \) forms a smooth manifold, this dimension coincides with its standard manifold dimension. For the general case, a rigorous treatment of \( \dim_{\mathbb{R}}(W_{\mathbb{R}}) \) can be found in \cite{realdim}.

For the real case, the following lemma is helpful.

\begin{lemma}[{\cite{Edidin, WangXu}}]
Let $W \subset \mathbb{C}^d$ be an algebraic variety. Then the real dimension of its real locus satisfies
\[
\dim_{\mathbb{R}}(W_{\mathbb{R}})\quad \leq\quad \dim(W).
\]
\end{lemma}

A particularly important class of algebraic varieties in our study consists of those \( W \subset \mathbb{C}^d \) satisfying \(\dim(W) = \dim_{\mathbb{R}}(W_{\mathbb{R}})\). A prototypical example is the variety \(\mathcal{M}_{d,r}\), which indeed exhibits this property.

\subsection{Admissible condition}
Let $W$ be the zero locus of a collection of homogeneous polynomials in $\C^{d}$ with $\dim( W)>0$ and let ${\ell_\alpha: \C^d\lra \C}$, $\alpha\in I$, be a family of (homogeneous) linear functions where $I$ is an index set that may be either finite or infinite.
\begin{definition} (see \cite{WangXu})  \label{defi-admissible}  
An algebraic variety $W\subset \C^d$ is called {\em admissible} with respect to the set $\{\ell_\alpha: \alpha \in I\}$ of homogeneous linear functions if 
\[
\dim \big(W\cap \{\vx\in \C^d:\ell_\alpha(\vx)=0\}\big) < \dim (W)
\]
holds for {\it every} $\alpha \in I$.
\end{definition}

The admissibility condition equivalently requires that no irreducible component of $W$ with dimension $\dim W$ can be entirely contained within any hyperplane $\{\vx\in \C^d: \ell_\alpha(\vx) = 0\}, \alpha\in I$ . This implies two  properties: 
\begin{enumerate}[label=(\roman*)]
    \item The intersection $W\cap \{\vx\in \C^d:\ell_\alpha(\vx)=0\}$ is always a proper algebraic subset of $W$ for every $\alpha \in I$.
    \item The dimension reduction 
    \[
    \dim(W\cap \{\vx\in \C^d:\ell_\alpha(\vx)=0\} ) = \dim W - 1
    \]
     holds  for every $\alpha \in I$.
\end{enumerate}

\begin{example}

To illustrate that the definition of admissibility is non-trivial, we present the following counterexample:
Let 
\[
W = \{X : X\in \mathbb{C}^{d\times d}, X=X^\top\}
\]
 be the set of symmetric matrices in $\mathbb{C}^{d\times d}$.
For any $Q\in \mathbb{C}^{d\times d}$, we define a linear function 
\[
\ell_Q: \mathbb{C}^{d\times d}\rightarrow \mathbb{C}
\]
 as $\ell_Q(X)=\text{Tr}(QX^\top)$ for $X\in \mathbb{C}^{d\times d}$, where Tr denotes the trace of a matrix.
We assert that $W$ fails to be admissible with respect to
\[
\{\ell_Q : Q\in \mathbb{C}^{d\times d}\setminus \{0\}\}.
\]
To prove this assertion, consider the matrix
\[
Q_0:=
\begin{pmatrix}
0 & 0 & \cdots & 0 & 1 \\
0 & 0 & \cdots & 0 & 0 \\
\vdots & \vdots & \ddots & \vdots & \vdots \\
0 & 0 & \cdots & 0 & 0 \\
-1 & 0 & \cdots & 0 & 0
\end{pmatrix}
\in \C^{d\times d}.
\]

For any $X\in W$, we have $\ell_{Q_0}(X)=0$, which follows from the symmetry of $X$ (i.e., $X=X^\top$).
Consequently,
\[
\dim (W\cap \{Y\in \C^{d\times d} :\ell_{Q_0}(Y)=0\}) {=}\dim W,
\]
violating the admissibility condition. Thus, $W$ is not admissible with respect to $\{\ell_Q : Q\in \mathbb{C}^{d\times d}\setminus \{0\}\}$.
This example demonstrates that the concept of admissibility is indeed substantive and not automatically satisfied for arbitrary algebraic variety  and set of linear functions.
\end{example}

\section{Optimal Sampling Number}

This section presents the findings from \cite{WangXu}, which address Problem \ref{que2} in most cases. These results can also be interpreted as applications of nonlinear algebra \cite{st}.
Specifically, this work establishes that for the equation $L_\vA(\mathbf{x}) = 0$, where $\vA \in \mathbb{C}^{m \times d}$ and $\vx \in W \subset \mathbb{C}^d$, the minimal number of measurements $m$ required to guarantee the uniqueness of the trivial solution $\mathbf{x} = 0$ is precisely $\dim (W)$.
This result is formally presented in the following theorem:

\begin{theorem}[\cite{WangXu}] \label{th:theo-LowRank} Let $W\subset \mathbb{C}^{d}$
  and $V_j\subset \mathbb{C}^d$, $j=1,\ldots,m$ be algebraic varieties. Assume that each $V_j$
  is admissible with respect to the set of linear functions $\{\ell_\mathbf{x} (\cdot):=L(\cdot,\mathbf{x}):~\mathbf{x}\in W\setminus \{0\}\}$.
   Define $V := V_1\times \dots \times V_m \subseteq (\mathbb{C}^{d})^m$. Then the following statements hold:
    \begin{enumerate}[label=(\roman*)]
     \item If $m \geq \dim (W)$, then there exists an algebraic subvariety $Z\subset V$ with $\dim(Z) \leq \dim(V) -\delta$ such that, for any $\vA=(\mathbf{a}_j)_{j=1}^m \in V\setminus Z$ and $\mathbf{x}\in W$,

$    L(\mathbf{a}_j,\mathbf{x}) = 0,\quad 1 \leq j\leq m, \text{ implies }  \mathbf{x}=0$. Here,  $\delta:=m-\dim (W)+1$.

\item If $m< \dim (W)$, then for any $\vA=(\mathbf{a}_j)_{j=1}^m \in V$, there exists a nonzero $\mathbf{x}_0 \in W$
  such that
$    L(\mathbf{a}_j, \mathbf{x}_0)=0 \text{ for all } 1 \leq  j \leq m.   $
\end{enumerate}
\end{theorem}

Theorem \ref{th:theo-LowRank} demonstrates that, for the complex case, the condition $m \geq \dim (W)$ is both necessary and sufficient to guarantee that the equation $L_\VA(\mathbf{x}) = 0$
  admits only the trivial solution $\mathbf{x} = 0$ within the set $W$.

We next use two examples to  elucidate Theorem \ref{th:theo-LowRank}.
\begin{example}
In Theorem \ref{th:theo-LowRank},  take $W = V_j = \mathbb{C}^d$, $L(\mathbf{a}_j, \mathbf{x}) := \langle\mathbf{a}_j, \mathbf{x}\rangle$, and $m = d$. In this scenario, ensuring that the homogeneous linear system 
\[
\langle\mathbf{a}_j, \mathbf{x}\rangle = 0,\qquad j = 1,\ldots,m,
\]
 admits only the trivial solution requires the matrix $\vA = (\mathbf{a}_j)_{j=1}^d \in \mathbb{C}^{d\times d}$ to be nonsingular. This condition is equivalent to $\vA = (\mathbf{a}_j)_{j=1}^d \in V \setminus Z$, where
$Z := \{\vA \in \mathbb{C}^{d \times d}: \det(\vA) = 0\}.$
 Notably, $\det(\vA)$ is a polynomial in the entries of $\vA$, rendering $Z$ an algebraic variety in $\mathbb{C}^{d \times d}$.
\end{example}

\begin{example}
 Compressed sensing has emerged as a particularly active discipline in recent years, primarily focused on the  recovery of sparse signals (see \cite{CS}). Let $k\leq d/2$. Set
\[
\Sigma_k:=\{\|\vz\|_0\leq k: \vz\in \C^d\},
\]
where $\|\vz\|_0$ denotes the number of non-zeros entries of $\vz$. A simple observation is that $\|\vx\|_0\leq k$ if and only if the product of any $k+1$ entries of $\vx$ is zero. Hence, $\Sigma_k$ is an algebraic variety in $\C^d$ with $\dim(\Sigma_k)=k$. 
We denote the measurement vectors as $\va_j \in \mathbb{C}^d$ for $j = 1, \ldots, m$. The measurements collected regarding $\vx$ are then expressed as:
$\innerp{\va_j, \vx},  j = 1, \ldots, m$.
A straightforward calculation reveals that $\vx - \vy \in \Sigma_{2k}$
  if $\vx, \vy \in \Sigma_k$. Therefore, one can uniquely recover any $\vx \in \Sigma_k$
  from $\innerp{\va_j, \vx}$, $j = 1, \ldots, m$ if and only if the following equations have only the trivial solution $\vx=0$:
\[
\innerp{\va_j,\vx}=0, \quad \vx\in \Sigma_{2k}.
\]
Applying Theorem \ref{th:theo-LowRank} with the choices $W:=\Sigma_{2k}$, $V_j:=\C^d$, and $L(\mathbf{a}_j, \mathbf{x}) := \langle\mathbf{a}_j, \mathbf{x}\rangle$ for $j=1,\ldots,m$, we deduce that the minimum number of measurements for the unique recovery of any $\vx\in \Sigma_k$ is $\dim(\Sigma_{2k})=2k$. 
Even though this result can be straightforwardly derived from linear algebra, we present this example to provide a concrete illustration of Theorem \ref{th:theo-LowRank}.
\end{example}

\begin{remark}
Theorem \ref{th:theo-LowRank} establishes the existence of an algebraic variety $Z \subset V$ such that $L_\VA(\mathbf{x}) = 0$
  admits only the trivial solution in $W$ when $\vA \in V \setminus Z$. An interesting open problem for future research is to explicitly characterize the ideal corresponding to $Z$.
\end{remark}

While Theorem \ref{th:theo-LowRank} primarily addresses the complex case, practical applications often require consideration of the real case. By combining Theorem \ref{th:theo-LowRank} with the assumption that $\dim_\mathbb{R} (V_\mathbb{R}) = \dim(V)$, we can extend our results to the real domain. The following corollary presents the corresponding result for the real case:

\begin{corollary}\label{co:real}(\cite{WangXu})
Let the conditions of Theorem \ref{th:theo-LowRank} be satisfied, and further suppose that ${\rm dim}_\R (V_\R) = {\rm dim}(V)$.
 Assume that  $m \geq \dim (W)$.  Then there exists a real algebraic subvariety  $\tilde{Z}\subset V_\R$ with ${\rm dim}_\R(\tilde{Z}) <
{\rm dim}_{\R}(V_\R)$ such that, for any $\vA= (\va_j )_{j=1}^m
\in V_\R \setminus  \tilde{Z}$ and $\vx\in W$, 
$L_\VA ( \vx) = 0$ implies $\vx=0$.
\end{corollary}

Corollary \ref{co:real} establishes that the condition $m \geq \dim (W)$ is sufficient to guarantee the existence of a matrix $\vA \in \mathbb{R}^{m \times d}$
  such that the equation $L_\vA(\mathbf{x}) = 0$
  admits only the trivial solution $\mathbf{x} = 0$ in $W$. However, it is important to note that this corollary does not assert the necessity of the condition $m \geq \dim W$. In fact, for the real case,  this condition is not necessary in general cases.

To utilize Theorem \ref{th:theo-LowRank} for determining the minimal number of measurements, two  steps are involved. First, it is necessary to calculate the dimension of an algebraic variety, which requires applying results from algebraic geometry. Second, it is essential to prove the admissibility property, often using techniques from matrix analysis.

We now apply the aforementioned results to investigate two  research problems in data science: phase retrieval and low-rank matrix recovery.

\section{phase retrieval}

The objective of phase retrieval is to reconstruct a vector $\mathbf{x} \in \mathbb{F}^d$
  from a set of quadratic measurements $y_j = \abs{\langle\mathbf{a}_j, \mathbf{x}\rangle}^2$, where $j = 1, \ldots, m$. In phase retrieval, a fundamental research problem is determining the minimal number of measurements $m$ required to recover any $\vx\in \F^d$
  from $(\abs{\langle\mathbf{a}_1, \mathbf{x}\rangle}^2,\ldots, \abs{\langle\mathbf{a}_m, \mathbf{x}\rangle}^2)$.
 A simple observation is that $\abs{\langle\mathbf{a}_j, \mathbf{x}\rangle}^2 = \abs{\langle\mathbf{a}_j, c \cdot \mathbf{x}\rangle}^2$
for any $c \in \F$ such that $\abs{c} = 1$. Consequently, in phase retrieval, one must acknowledge that the recovery of $\vx$ is only possible up to a unimodular constant.

  At first glance, the measurements appear nonlinear, and $\vx$ belongs to the entire space, seemingly unrelated to algebraic varieties. However, one can employ the PhaseLift method to transform these measurements into linear ones, at the cost of the object now belonging to an algebraic variety in a higher-dimensional space.

We shall now elucidate this concept in  detail. An observation is that
\begin{equation}\label{eq:lift}
y_j=\abs{\langle\mathbf{a}_j, \mathbf{x}\rangle}^2=\innerp{\va_j\va_j^*, \vx\vx^*}, \quad j=1,\ldots,m.
\end{equation}
In this context,  we define the inner product as $\innerp{\vA,X}=\Tr(\vA X^*)$ for matrices $\vA, X\in \F^{d\times d}$.
We define $X^*$ as the Hermitian (or conjugate) transpose if $\F=\C$, and the standard transpose if $\F=\R$.

Let $\cS_{d,r}(\F)$ denote the set of (Hermitian) Symmetric matrices with rank less than or equal to $r$:
\[
\cS_{d,r}(\F)\,\,:=\,\, \{Q\in \F^{d\times d}: Q=Q^*, {\rm rank}(Q)\leq r\}.
\]
We also set
\[
\cM_{d,r}({\mathbb F})\,\,:=\,\, \{Q\in {\mathbb F}^{d\times d}: {\rm rank}(Q)\leq r\}.
\]
Here, ${\mathbb F}\in \{\mathbb{R}, \mathbb{C} \}$. The explicit inclusion of ${\mathbb F}$ in the notations $\cS_{d,r}(\F)$ and $\cM_{d,r}({\mathbb F})$  serves to specify whether the matrices are real or complex, thereby emphasizing their nature.
For the sake of brevity, we frequently omit the field $\F$ in the notation $\cS_{d,r}(\F)$ and $\cM_{d,r}({\mathbb F})$, as its specification (whether $\R$ or $\C$) can be inferred from the context.

Consequently, we can reformulate the generalized phase retrieval problem as follows:
{\em
To recover $X\in \cS_{d,1}$ from linear measurements
\[
y_j=\innerp{\vA_j, X},\qquad j=1,\ldots,m,
\]
where $\vA_j\in V_j:=\cS_{d,r_j}$ are given matrices for $j=1,\ldots,m$, and $r_j\in [1,d]$.}

 Here, we have removed the requirement for the matrix $\vA_j$ to be of rank one, and instead consider a given rank $r_j$. 
 When $r_j, j=1,\ldots,m$
  are set to 1, the problem simplifies to the phase retrieval case discussed earlier.

\subsection{Real case} 
We begin by presenting the results for the  case where $\F=\R$. Let $\A=\{\vA_j\}_{j=1}^m$
  denote the measurement matrix set, with $A_j\in \cS_{d,r_j}$. 
  We say that $\A$ possesses the {\em phase retrieval property} in $\R^d$ if, for any $\vx,\vy \in \R^d$, the condition $\vx^\top \vA_j\vx=\vy^\top \vA_j\vy$ for all $j=1,\ldots,m$ implies that $\vx=\pm \vy$.
    In this subsection, we focus on the question: {\em How many matrices does ${\mathcal A}$ need to have so that ${\mathcal A}$ has the phase retrieval property in $\R^d$?}
    A simple calculation shows that 
  \[
\vx^\top \vA_j\vx - \vy^\top \vA_j\vy = \innerp{\vA_j, {Q}},
\]
where
\begin{equation*}
{Q}=\frac{1}{4}(\vx-\vy)(\vx+\vy)^\top\in {\mathcal M}_{d,1}{(\R)}.
\end{equation*}
Therefore, if $\A$ possesses the phase retrieval property in $\R^d$, then the equation
\[
 X \in \mathcal M_{d,1}{(\R)}, \quad \innerp{\vA_j, X} = 0,\,\, j=1,\ldots,m
\]
 admits only the trivial solution $X = \mathbf{0}$. Recall that ${\rm dim}({\mathcal M}_{d,1})=2d-1$. Then we can employ Corollary \ref{co:real}
  to obtain that
  \begin{theorem}(\cite{WangXu}) \label{theo-RealPRgeneric}
   Let $m\geq 2d-1$ and $1 \leq r_1, \dots, r_m \leq d$. Then a generic $\A=\{\vA_j : \vA_j\in \cS_{d,r_j}(\R), j=1,\ldots,m\} $  has the phase retrieval property in $\R^d$.
\end{theorem}
Next, we outline the proof of Theorem \ref{theo-RealPRgeneric}, using this example to illustrate  how to establish the admissible condition (as defined in Definition \ref{defi-admissible}).
\begin{proof}[Sketch of proof of Theorem \ref{theo-RealPRgeneric}]

In Corollary \ref{co:real}, let $V = V_{r_1} \times \cdots\times V_{r_N}$, where each $V_{r_j}$ denotes the set of complex symmetric matrices in $\mathbb{C}^{d\times d}$ with rank at most $r_j$. Each $V_{r_j}$ is an algebraic variety defined by the zero locus of a set of homogeneous polynomials.
A simple observation is that $(V_{r_j})_\R=\cS_{d,r_j}(\R)$.
 It is well known that $\dim(V_{r_j}) = dr_j - \frac{r_j(r_j-1)}{2}$, and its real dimension $\dim_\mathbb{R}((V_{r_j})_\mathbb{R})$ is also equal to this value. Consequently, $\dim(V) = \dim_\mathbb{R}(V_\mathbb{R})$.

We define $W = {\mathcal M}_{d,1}(\mathbb{C})$ and the linear function $L(\mathbf{A},Q) = \langle\mathbf{A}, Q\rangle$ with $\vA, Q\in \C^{d\times d}$. If we assume that each $V_{r_j}$ is admissible with respect to the set of functions $\{L(\cdot, Q) \text{ for } Q \in W\setminus\{0\}\}$, then our theorem follows immediately from Corollary \ref{co:real}.

Thus, it remains only to demonstrate the admissibility of $V_{r_j}$. To this end, it suffices to show that for a generic point $\vA_0 \in V_{r_j}$ and any nonzero rank-one matrix $Q_0=\vx_0\vy_0^\top\in W$ with $\vx_0, \vy_0 \in \mathbb{C}^d$, the function $L(\cdot,Q_0) \not\equiv 0 $  in any small neighborhood of $\vA_0$ within $V_{r_j}$.
If $L(\vA_0, Q_0) \neq 0$, the condition is immediately satisfied. Therefore, assume $L(\vA_0, Q_0) = 0$.
Applying the Takagi factorization to $\vA_0$, we write it as:
$$
      \vA_0 = \sum_{i=1}^s \vz_i \vz_i^\top,
$$
where $s\leq r_j, \vz_i\in \C^d, i=1,\ldots,s$.

Next, we consider a perturbation of $\vA_0$ by defining $\hat\vz_1=\vz_1 +t\vu$ for some $\vu\in\C^{d}$ (to be chosen later). We then define $\vA_\vu = \hat\vz_1\hat\vz_1^\top+\sum_{i=2}^s \vz_i \vz_i^\top$.

Then, substituting $\vA_\vu$ and using our assumption that $\innerp{\vA,Q_0}=0$, we compute $\innerp{\vA_0,Q_0}$ as:
$$
     \innerp{\vA_\vu,Q_0}= t^2 (\vy_0^\top\vu)(\vu^\top\vx_0) + t (\vy_0^\top\vu + \vu^\top\vx_0).
$$
Since $\vu$ is an arbitrary vector, we can always choose it such that $(\vy_0^\top\vu)(\vu^\top\vx_0)\neq 0$. With such a choice of $\vu$, for sufficiently small nonzero $t$, the expression for $\innerp{\vA_\vu,Q_0}\neq 0$, as it is dominated by the lowest-order non-zero term in $t$. This demonstrates that $L(\cdot,Q_0) $ is not identically zero in any small neighborhood of $\vA_0$ within $V_{r_j}$.
\end{proof}

In the case where $r_1=\cdots=r_m=1$, that is, when $\vA_j=\va_j\va_j^\top$ with $\va_j\in \R^d$ for $j=1,\ldots,m$, the lower bound of $2d-1$ is tight (see \cite{BCE06}). Indeed, Balan, Casazza, and Edidin present the necessary and sufficient conditions under which the set $\{\va_j\va_j^\top\}_{j=1}^m$ possesses the phase retrieval property in $\R^d$.

\begin{theorem}\label{th:rankone}(\cite{BCE06})
Suppose that $\{\va_1,\ldots,\va_m\}\subset \R^d$. 
If $\{\va_j\va_j^\top\}_{j=1}^m$ has phase retrieval property in $\R^d$ then  
for every subset $S\subset \{1,\ldots,m\}$, either {${\rm span}\{\va_j: j\in S\}=\R^d$ }or {${\rm span}\{\va_j: j\in S^c\}=\R^d$}.
\end{theorem}

In fact,  for rank-one case, Theorem \ref{th:rankone} characterizes an algebraic subvariety ${\tilde Z}\subset (\cS_{d,1}\times \cdots \times\cS_{d,1})_\R$ such that $\A=\{\va_j\va_j^\top\}_{j=1}^m$ has the phase retrieval property if and only if $\A$ does not belong to  ${\tilde Z}$.

Without the constraint of rank one of $\vA_j, j=1,\ldots,m$, the lower bound of $2d-1$ presented in Theorem \ref{theo-RealPRgeneric} is not generally tight. To elaborate on this, we define:
\begin{equation}\label{eq:mrd}
   \m_\R(d):=\min\{m: \text{there exists  a phase retrievable $\{\vA_j: \vA_j\in \cS_{d,d}(\R), j=1,\ldots,m\}$ in $\R^d$} \}.
\end{equation}

The following theorem demonstrates that $\m_\R(d) \leq 2d-2$
  when $d$ is even. This result implies that the lower bound of ${\rm dim}({\mathcal M}_{d,1}) = 2d-1$
  presented in Theorem \ref{theo-RealPRgeneric} is not tight for even values of $d$.

\begin{theorem}\cite{WangXu} \label{th:rmin}
Let $\m_\R(d)$  be defined in (\ref{eq:mrd}). Then we have
\begin{enumerate}
\item $\m_\R(d) \leq 2d-1$  any odd $d$ and $\m_\R(d) \leq 2d-2$ for any even $d$.
\item For any $k\geq 1$,
$$
   \m_\R(d)= \left\{\begin{array}{cl} 2d-1,  & ~d=2^k+1,\\
   2d-2, & ~d=2^k+2. \end{array}\right.
$$
\item For any $d\geq 5$,
$$
   \m_\R(d) \geq \left\{\begin{array}{cl} 2d-6\lfloor\log_2(d-1)\rfloor+6, &~~\mbox{$d$ odd},\\
                2d - 6\lfloor\log_2(d-2)\rfloor+4, &~~\mbox{$d$ even}.
                \end{array}\right.
$$
\end{enumerate}
\end{theorem}

We present the following open problems related to generalized phase retrieval in real case:

\begin{problem}
 Determine the exact value of $\m_\R(d)$  for all integers $d \geq 2$.
 \end{problem}
 
 \begin{problem}
Given the conditions specified in Theorem \ref{theo-RealPRgeneric}, characterize the set $\A = \{\vA_j\}_{j=1}^m$
 for which $\A$ possesses the phase retrieval property in $\R^d$.
 \end{problem}

  \subsection{Complex case}
We now turn our attention to the complex case, namely when $\F=\C$.

\subsubsection{Standard phase retrieval}
  We begin from standard phase retrieval case which aims to recover any $\vx\in \C^d$ from $y_j=\abs{\innerp{\va_j,\vx}}^2, j=1,\ldots,m$. Here, $\va_j\in \C^d$ are known measurements vectors. 
  We say that $\vA=(\va_1,\ldots,\va_m)\in \C^{d\times m}$ possesses the {\em phase retrieval property} in $\C^d$ if
    $   \abs{\innerp{\va_j,\vx}}^2=\abs{\innerp{\va_j,\vy}}^2, \text{for  } j=1,\ldots,m,$   we have $\vx=c\cdot \vy$ with $c\in \C, \abs{c}=1$ for any $\vx,\vy \in \C^d$.

For the complex case, the same question remains open for standard phase retrieval:
\begin{center}
{\em What is the minimum number of vectors required in the set $\vA$ to ensure it possesses the phase retrieval property in $\mathbb{C}^d$?}
\end{center}
 It is established that in the standard phase retrieval setting, $m \geq 4d-4$ generic vectors $\{\va_j\}_{j=1}^m \subset \C^d$  possess the phase retrieval property \cite{BCMN, CEHV15}. Furthermore, $m = 4d-4$ is minimal when $d = 2^k + 1$, where $k \geq 1$ \cite{CEHV15}. Vinzant \cite{V15} constructed an example in $\C^4$  with $m = 11 = 4d-5 < 4d-4$ vectors $\{\va_j\}_{j=1}^{11}$
  such that $ (\va_j)_{j=1}^{11}$ is phase retrievable.  The result implies that $m = 4d-4$ is not minimal for some dimensions $d$ in standard phase retrieval. To date, the smallest $m$ remains unknown, even for $d = 4$.
In the opposite direction, in \cite{HMW13}, Heinosaari-Mazzarella-Wolf  established a lower bound of $m \geq 4d-3-2\alpha$ for the minimal number of measurements, where $\alpha$ denotes the number of 1's in the binary expansion of $d-1$. 

\subsubsection{Generalized phase retrieval}

Let 
\[
\A:=\{\vA_j: \vA_j\in \mathcal{S}_{d,r_j}, j=1,\ldots,m\}
\]
 be the set of measurement matrices, where $r_j\in [1,d]$
  is a given integer for each $j$.
Analogous to the real case, the aim of generalized phase retrieval is to recover $\vx\in \C^d$  from 
\[
y_j=\vx^*\vA_j\vx, j=1,\ldots,m.
\]
As in the previous case, we say that the set $\A$
possesses the {\em phase retrieval property in $\mathbb{C}^d$ } if, for any $\vx,\vy \in \mathbb{C}^d$
  satisfying $\vx^*\vA_j\vx=\vy^*\vA_j\vy \text{ for all } j=1,\ldots,m,$ we have $\vx=c\vy$ for some $c\in \mathbb{C}$ with $\abs{c}=1$.
A simple calculation shows that 
$
\vx^*\vA_j\vx=\vy^*\vA_j\vy
$ 
if and only if 
$\Tr(\vA_j(\vx\vx^*-\vy\vy^*))=0$. Hence, we have
\begin{lemma}
Let $\A=\{\vA_j: \vA_j\in \mathcal{S}_{d,r_j}, j=1,\ldots,m\}$ be the set of measurement matrices, where $r_j\in [1,d]$
  is a given integer for each $j$. If the system of equations
 \[ 
X\in \mathcal{S}_{d,2}, \quad \text{Tr}(\vA_jX)=0, \quad j=1,\ldots,m
\]
 admits only the zero solution, then $\A$ possesses the phase retrieval property in $\mathbb{C}^d$.
\end{lemma}

Recall that $\mathcal{S}_{d,r_j}$  denotes the set of $d\times d$ Hermitian matrices  with rank less than or equal to $r_j$.
Note that the operation of conjugation is not an algebraic operation and hence $\cS_{d,r_j}$ is {\em not} an algebraic variety in complex space.
To overcome this problem, we introduce the  linear map $\tau: \C^{d\times d} \rightarrow \C^{d\times d}$ which is defined by 
$$
    \tau(\vA) = \frac{1}{2} (\vA+\vA^\top) +\frac{i}{2} (\vA-\vA^\top).
$$ 

A straightforward calculation demonstrates that $\tau$ is an isomorphism on $\mathbb{C}^{d\times d}$. Moreover, when restricted to $\mathbb{R}^{d\times d}$, $\tau$ becomes an isomorphism from $\mathbb{R}^{d\times d}$  to the set of Hermitian matrices, which is precisely $\mathcal{S}_{d,d}$
  as previously defined.
  Set $L(\tilde{\vA}_j,X):=\inner{\tau(\tilde{\vA}_j),X}$. 
   Using $\tau$ and $L(\tilde{\vA}_j,x)$, we can take
  \[
  \tilde{\vA}_j\in (V_j)_\R=\{\vA\in \R^{d\times d}: {\rm rank} (\tau(\vA))\leq r_j\}.
  \]
  Taking  $W:= \{X\in \C^{d\times d}: \rank(X)\leq 2\}$ in Corollary  \ref{co:real}, we obtain the following result:

\begin{theorem}(\cite{WangXu})\label{theo-ComplexPRgeneric}
   Let $m\geq 4d-4$ and $1 \leq r_1, \dots, r_m \leq d$.
Let $\A=\{\vA_j: \vA_j\in \mathcal{S}_{d,r_j}(\C), j=1,\ldots,m\}$ be the set of measurement matrices. 
 Then a generic $\A$  has the phase retrieval property in $\C^d$.
\end{theorem}

Although complex cases are considered here, as we require the measurement matrices $\vA_j, j=1,\ldots,m$ to be Hermitian, we  use the transformation $\tau$ to convert them to the real domain.
Hence, in the proof of Theorem \ref{theo-ComplexPRgeneric}, we take $\vA_j=\tau(\tilde{\vA}_j)$, where $\tilde{\vA}_j$ is a {\em real} matrix, and $L(\tilde{\vA}_j,X)=\innerp{\tau(\tilde{\vA}_j),x}=\innerp{\vA_j,X}$.
Therefore, we ultimately still need to deal with the real algebraic variety situation. This also means that the bound of $4d-4$ given by the above theorem is not always tight.
To examine the minimal measurement number,
we set
\begin{equation}\label{eq:cmrd}
   \m_\C(d):=\min\{m: \text{there exists  a phase retrievable $\{\vA_j: \vA_j\in \cS_{d,d}, j=1,\ldots,m\}$ in $\C^d$} \}.
\end{equation}
Then we have
\begin{theorem}(\cite{WangXu})\label{th:clower}
Let $d>4$. Then $4d-2-2\alpha+\epsilon_\alpha \leq \m_\C(d)\leq 4d-3-\alpha-\delta$,
where $\alpha = \alpha(d-1)$ denotes the number of $1$'s in the binary expansion of $d-1$,
$$
\epsilon_\alpha= \left\{\begin{array}{cl} 2  & ~d \text{ odd},\, \alpha \equiv 3 \wmod 4,\\
   1 &   ~d \text{ odd}, \,\alpha \equiv 2 \wmod 4,\\
   0  & ~\text{otherwise.}
   \end{array}\right.
   \shsp \mbox{and} \shsp
   \delta =  \left\{\begin{array}{cl} 0  & ~d \text{ odd},\\
   1 &   ~d \text{ even}.
   \end{array}\right.
$$
\end{theorem}
\begin{remark}
The proof of Theorem \ref{th:clower} employs the  topological results on the embedding of projective spaces into Euclidean space (see \cite{WangXu}).
Indeed, if ${\mathcal A}=\{\vA_j: \vA_j\in \cS_{d,d}, j=1,\ldots,m\}$   is phase retrievable  in $\C^d$, then it can produce an embedding of ${\mathbb P}(\C^d)$ in $\R^{m-1}$.
 In topology, numerous lower bounds for $m$ have been established to guarantee the embeddability of ${\mathbb P}(\C^d)$ in $\R^{m-1}$ (see \cite{embd}).
  \end{remark}
  Based on Theorem \ref{th:clower}, we are able to precisely determine the values of $\m_\C(d)$ for certain integer values of $d$.
\begin{theorem}(\cite{WangXu}) \label{th:cexact}
 We have the following exact values for $\m_\C(d)$:
$$
   \m_\C(d)= \left\{\begin{array}{cl} 4d-4  & ~d=2^k+1, \,k>1,\\
   4d-6 & ~d=2^k+2, \,k>1, \\ 
    4d-5 & ~d=2^k+2^j+1, \,k>j>0,\\
    4d-6 & ~d=2^k+2^j+2^l+1, \,k>j>l>0.\end{array}\right.
$$
Also, $\m_\C(2)=3$.
\end{theorem}

We would like to present the following open problem related to generalized phase retrieval in the complex case, building on the issue previously addressed in the real case:

\begin{problem}
 Determine the exact value of $\m_\C(d)$  for all integers $d \geq 2$.
 \end{problem}

\section{Low-rank Matrix recovery}

The aim of low-rank matrix recovery is to recover any 
 $Q\in \cM_{d,r}({\mathbb F})$ from the linear measurements 
\begin{equation*}
y_j:=\innerp{\vA_j,Q}, \quad j=1,\ldots,m,
\end{equation*}
where $\vA_j\in \cM_{d\times d}({\mathbb F}), j=1,\ldots, m$ are known measurement matrices.
Throughout the rest of this paper, we assume that $r\leq d/2$.
One aims to leverage the prior information that the matrix $Q$ has a rank not exceeding $r$ in order to reconstruct $Q$ using fewer linear measurements, specifically $m \ll d^2$. The matrix $Q$ is one solution to the following underdetermined linear system with the unknown variable $X \in \mathbb{F}^{d \times d}$:
\begin{equation}\label{eq:unde}
y_j = \innerp{ \vA_j, X }, \quad j=1,\ldots,m.
\end{equation}

Based on the assumption that $\text{rank}(Q) \leq r$, we have
\[
Q \in \{X \in \mathbb{F}^{d \times d} : y_j = \langle \vA_j, X \rangle, \quad j=1,\ldots,m\} \cap \mathcal{M}_{d,r}({\mathbb F}).
\]
For convenience, we set 
\[
 \MM_\A(Q):=(\innerp{\vA_1,Q},\ldots,\innerp{\vA_m,Q})\in \F^m
 \]
 where $\A=\{\vA_j\}_{j=1}^m$.
 We say that $\MM_\A$ is {\em injective} on $\mathcal{M}_{d,r}({\mathbb F})$ if and only if, for any $Q_1, Q_2 \in \mathcal{M}_{d,r}({\mathbb F})$, the condition $\MM_\A(Q_1) = \MM_\A(Q_2)$
implies $Q_1 = Q_2$.

A popular programming for recovering $Q$ is 
\begin{equation}\label{eq:zuixiao}
\begin{aligned}
&\argmin\limits_{\tiny X\in {\mathbb F}^{d\times d}} \quad{\rm rank} (X)\\
&{\rm s.t.  }\quad  \innerp{\vA_j,X}= \innerp{\vA_j,Q},\quad j=1,\ldots,m.
\end{aligned}
\end{equation}

We use the following lemma to demonstrate the equivalence of these statements.
\begin{lemma}
Assume that $\A=\{\vA_j\}_{j=1}^m\subset \cM_{d,d}({\mathbb F})$ are measurement matrices. 
The followings are equivalent:
\begin{enumerate}
\item 
For any $Q\in \cM_{d,r}({\mathbb F})$, we have
\[
\{Q\}= \{X \in \mathbb{F}^{d \times d} : y_j = \langle \vA_j, X \rangle, \quad j=1,\ldots,m\} \cap \mathcal{M}_{d,r}({\mathbb F}).
\]
\item 
The $\MM_\A$ is injective on $\mathcal{M}_{d,r}({\mathbb F})$.
\item For any $Q\in \cM_{d,r}({\mathbb F})$, the solution to (\ref{eq:zuixiao}) is $Q$.
\end{enumerate}
\end{lemma}

Now we state the  fundamental problem in low-rank matrix recovery as follows:
\begin{problem}
What is the minimum number of measurements $m$ such that there exist $m$ matrices $\mathcal{A} = \{\vA_j\}_{j=1}^m$
 for which $\MM_\A$  is injective on $\mathcal{M}_{d,r}({\mathbb F})$?
\end{problem}

The following lemma proves useful for our analysis.
\begin{lemma}\label{le:uniq}(\cite{uniq})
Suppose that $r\leq d/2$. The map ${\bf M}_{\mathcal A}$ is {not} injective on $\ML_{d,r}(\H)$ if and only if there is a nonzero $Q\in \ML_{d,2r}(\H)$ for which
\[
{\bf M}_{\mathcal A}(Q)=0.
\]
\end{lemma}

Based on Lemma \ref{le:uniq}, it was established in \cite{uniq} that for \( m \geq 4dr - 4r^2 \),  Gaussian random matrices \( \mathcal{A} = \{\vA_j\}_{j=1}^m \) ensure that \( \MM_\A \) is injective on \( \mathcal{M}_{d,r}({\mathbb F}) \) with probability 1. Furthermore, they conjectured that \( 4dr - 4r^2 \) is a tight bound, meaning that if \( m < 4dr - 4r^2 \), then for any \( \mathcal{A} = \{\vA_j\}_{j=1}^m \), \( \MM_\A \) is not injective on \( \mathcal{M}_{d,r} \).
We next confirm this conjecture for the complex case.

For ${\mathbb F}={\mathbb C}$, by  taking $W={\mathcal M}_{d,2r}(\C)$ in Theorem \ref{th:theo-LowRank}, we  obtain the following theorem:

\begin{theorem}(\cite{Xu15})\label{th:comp}
Suppose that $r\leq d/2$.
Consider $m$ matrices ${\mathcal A}=\{\vA_1,\ldots,\vA_m\}\subset \C^{d\times d}$ and the mapping ${\bf M}_{\mathcal A}:\C^{d\times d}\rightarrow \C^m$. The following holds

(a) If $m\geq {\rm dim}({\mathcal M}_{d,2r}(\C))= 4dr-4r^2$ then ${\bf M}_{\mathcal A}$ is injective on $\ML_{d,r}(\C)$ for generic matrices $\vA_1,\ldots,\vA_m$.

(b) If $m<4dr-4r^2$, then ${\bf M}_{\mathcal A}$ is not injective on $\ML_{d,r}(\C)$.

\end{theorem}

\begin{remark}
By the $m$ generic matrices in $\H^{d\times d}$  we mean  $[\vA_1,\ldots,\vA_m]$ corresponds to a point in a non-empty  Zariski open subset
of $\H^{md^2}$ which is also open and dense in the Euclidean topology.
\end{remark}

Theorem \ref{th:comp} shows that, for the complex case, the minimal measurement number for recovering any $Q\in {\mathcal M}_{d,r}(\C)$ is $4dr-4r^2$ provided $r\leq d/2$.

For the real case,  we introduce the following result:

\begin{theorem}(\cite{Xu15})\label{th:rcomp}
Suppose that $r\leq d/2$.
Consider $m$ matrices ${\mathcal A}=\{\vA_1,\ldots,\vA_m\}\subset \R^{d\times d}$ and the mapping ${\bf M}_{\mathcal A}:\R^{d\times d}\rightarrow \R^m$. The following holds

(a) If $m\geq 4dr-4r^2$ then ${\bf M}_{\mathcal A}$ is injective on $\ML_{d,r}(\R)$ for generic matrices $\vA_1,\ldots,\vA_m$.

(b) 
Suppose that $d=2^k+r, k\in\Z_+,$ or $d=2r+1$. If $m<4dr-4r^2$, then ${\bf M}_{\mathcal A}$ is  not injective on $\ML_{d,r}(\R)$.

\end{theorem}

For the real case, Theorem \ref{th:rcomp} establishes the tightness of the bound $4dr-4r^2$ when $d$ takes the form $2^k+r$ or $2r+1$. A natural question then arises as to whether this bound is generally tight. However, in \cite{Xu15}, a counterexample demonstrates that it is not universally applicable.
 Specifically, 
 for the specific case where $(d,r)=(4,1)$, there exist $11 = 4d-5$ matrices ${\mathcal A}=\{\vA_1,\ldots,\vA_{11}\}\subset \mathbb{R}^{4\times 4}$ such that ${\bf M}_{\mathcal A}$ is injective on ${\ML}_{4,1}(\mathbb{R})\subset \mathbb{R}^{4\times 4}$. 
The problem of constructing these $11$ matrices reduces to ensuring that a specific polynomial system (defined by 16 variables and 27 equations—11 linear and 16 cubic) admits no non-zero real solutions (see \cite{Xu15} for details).
The 11 matrices, obtained via computer random search in \cite{Xu15}, are listed as follows:
{
\begin{equation*}\label{eq:11matr}
  \begin{aligned}
A_1&= \begin{pmatrix} -4 & 1 & 3& 4 \\
 -4 & 4 & 4 & 3\\
 4 & -3 & 0 & -3 \\
  0 & -4 & 2 & 1 \\
 \end{pmatrix},
  A_2=\begin{pmatrix}
   0 & 3 & -1& -1 \\
 0 & -2 & -1 & 2\\
 0 & 3 & -2 & 3 \\
  1 & -1 & -3 & 2 \\
 \end{pmatrix},
 A_3= \begin{pmatrix}
   -1 & -4 & -1& -1 \\
 4 & 0 & -1 & 1\\
 -2 & 0 & 0 & 2 \\
  0 & -1 & 2 & 2 \\
 \end{pmatrix}\\
  A_4&= \begin{pmatrix}
   -2 & -2 & 4& 1 \\
 -2 & 0 & 2 & 3\\
 1 & -2 & -4 & 3 \\
  -3 & 3 & 4 & -2 \\
 \end{pmatrix},
  A_5=  \begin{pmatrix}
  4  & 2 & -4& -4 \\
 -4 & -3 & 0 & 0\\
 1 & -4 & 4 & -2 \\
  3 & 0 & 2 & 0 \\
 \end{pmatrix},
A_6= \begin{pmatrix}
 2  & 2 & 3& 4 \\
 2 & -4 & 3 & 1\\
 0 & -2 & 1 & -2 \\
  -1 & 0 & -1 & -4 \\
 \end{pmatrix}\\
 A_7&=\begin{pmatrix}
 2  & 1 & 4& 0 \\
-1 & -3 & 0 & -1\\
 4 & -1 & -4 & 3 \\
 0 & 3 & 0 & 4 \\
 \end{pmatrix},
 A_8=\begin{pmatrix}
 0  & 3 & -1& 2 \\
4 & 2 & 1 & 1\\
 -2 & -1 & 3 & 4 \\
 3 & 0 & 3 & 3 \\
 \end{pmatrix},
  A_9=\begin{pmatrix}
2  & -1 & 4& -4 \\
-2 & 2 & 3 & -1\\
 -1 & 1 & 4 & -1 \\
 -3 & -4 & 4 & 3 \\
 \end{pmatrix}\\
  A_{10}&= \begin{pmatrix}
-4  & 2 & 0& -1 \\
4 & 1 & 0 & 4\\
 -1 & -3 & 4 & 1 \\
 -3 & 2 & 4 & -4 \\
 \end{pmatrix},
   A_{11}= \begin{pmatrix}
1  & 1 & -2& 0 \\
3 & 0 & -2 & -4\\
 2 & -4 & -2 & 4 \\
 4 & 3 & 2 & -2 \\
 \end{pmatrix}.
 \end{aligned}
\end{equation*}}

\begin{remark}
Notably, for the real case where $r \leq d/2$, determining the minimal number of linear measurements $m$ required to recover any matrix $Q \in \mathcal{M}_{d,r}(\mathbb{R})$ is still an open problem.
\end{remark}

\begin{remark}
In the field of low-rank matrix recovery, the concept of almost everywhere low-rank matrix recovery is also of interest. This refers to the ability to recover all rank $r$ matrices, except for a set of measure zero. Rong-Wang-Xu \cite{almost} apply the algebraic variety method to investigate this almost everywhere recovery.
\end{remark}

\section{Summary} This paper introduces the algebraic variety method, an approach for quantifying the minimal number of measurements needed to recover signals constrained to an algebraic variety. We demonstrate its applicability to well-known problems, such as phase retrieval and low-rank matrix recovery. Throughout the discussion, we identify several open problems.

Our current findings primarily focus on the complex and real domains; however, extending these results to the quaternion domain presents an intriguing avenue for further exploration. Additionally, while this work emphasizes determining the minimal number of measurements required, we have also dedicated efforts to investigating the performance of various models and algorithms, particularly in the context of phase retrieval (e.g., \cite{XX24,XXX24,HX24,GX17}). An extension of our study would involve applying these performance analyses to a broader context, specifically concerning signals defined on arbitrary algebraic varieties.

\end{document}